\documentclass[10pt,a4paper,oneside]{article}
\usepackage[utf8]{inputenc}
\usepackage[english]{babel}
\usepackage{amsmath}
\usepackage{amsfonts}
\usepackage{graphics}
\usepackage{graphicx}
\usepackage[affil-it]{authblk}
\usepackage{amssymb}
\usepackage{epstopdf}
\usepackage[hmargin=2cm,vmargin=3cm]{geometry}
\usepackage{setspace}
\DeclareGraphicsExtensions{.pdf,.jpg,.png}
\title{A branching network model for T cell dissemination in adaptive immune response}
\date{}
\author{A. Boianelli$^{1}$ and A. Vicino$^{2}$}
\affil{$^{1}$ Systems Medicine of Infectious Diseases Group, Department of Systems Immunology, Helmholtz Centre for Infection Research, Inhoffenstrasse 7, Braunschweig, Germany  }
\affil{$^{2}$Dipartimento di Ingegneria dell'Informazione e Science Matematiche, Universit\`a di Siena, Via Roma 56 , 53100 Siena, Italy.}
\begin{document}
\maketitle

\section*{Abstract}
In this paper we consider a model based on branching process theory for the proliferation and the dissemination network  of T cells in the adaptive immune response. A multi-type Galton Watson branching process is assumed as the basic proliferation mechanism, associated to the migration of T cells of the different generations from the draining lymph node to the spleen and other lymphoid organs. Time recursion equations for the mean values and the covariance matrices of the the cell population counts are derived in all the compartments of the network model. Moreover, a normal approximation of the log-likelihood function of the cell relative frequencies is derived, which allows one to obtain estimates of both the probability parameters of the branching process and the migration rates in the various compartments of the network.
\section{Moment computation for the network compartments}
In this Section we derive the first and second order moments of cell counts in the different compartments of the network model. A  proliferation mechanism following a memoryless branching process known in the literature as a multi-type Galton Watson process is adopted. Population mean values and covariance matrices are essential tools both for stochastic simulation of the network and for implementing the parameter estimation procedure described in Section {\bf S2}.
We start our development by considering the source node of our network, i.e., the draining lymph node.

\subsection*{Draining lymph node}

The classical multi-type Galton Watson branching process (MGW)is a prototypical branching process, representing  the evolution  of a population whose members reproduce and die subject to random laws (see the classical textbooks ({\it 1, 2\/})). The generation of a cell, which is defined as the number division steps the cells undergoes before its birth, represents the type of that cell. We will denote by  $\Delta t$  the sampling time (or time step) of the discrete time process describing the proliferation. The generic time point $t=n \ \Delta t$ will be denoted by $n$. Let the state of the population in the draining lymph node at time $n$ be represented by the vector $\mathbf{Z}_{dr}(n) = [Z_{dr,0}(n), Z_{dr,1}(n), Z_{dr,2}(n), \ldots, Z_{dr,p}(n)]$, where $Z_{dr,i}(n)$ are discrete random variables given by the counts of cells after $i$ divisions, being $p$ the maximum generation considered.\\
It is well known that a MGW process is a homogeneous vector Markov process
$\mathbf{Z}_{dr}(n)$, $ n \in \mathbb{N}$, with the following properties  (see ({\it 1, 2\/})):
 \begin{enumerate}
 \item[P1)]each cell action is independent from others;
 \item[P2)]each cell offspring generates its own branching process;
 \item[P3)]at each time point, the cell has no memory about previous time steps.
 \end{enumerate}

 In the most simple setting, a cell of type   $i$
 can make three probabilistic decisions during each time step $\Delta t$:
 \begin{enumerate}
\item[D1)]
remain in the same generation $i$ with probability $\delta_{i}$;
\item[D2)]
divide to generate two cells of type $i + 1$ with probability $\gamma_{i}$;
\item[D3)]
die with probability $\alpha_{i}$.
\end{enumerate}
The transition from $\mathbf{Z}_{dr}(n)$ to  $\mathbf{Z}_{dr}(n+1)$ in one time step is regulated by the probability generating function (pgf):

\begin{equation}
\mathbf{f}(\mathbf{s})=[f_{0}(\mathbf{s}),f_{1}(\mathbf{s}), f_{2}(\mathbf{s}),\ldots, f_{p}(\mathbf{s})]^T \ ,
\end{equation}
where
$$
\mathbf{s}=[s_{0}, s_{1}, s_{2},\dots, s_{p}] \in \mathbb{C}^{p+1}, \vert s_i \vert \leq 1 \ ,
$$
and
\begin{equation}
f_i(\mathbf{s})=\delta_{i}s_{i}+\gamma_{i}s_{i+1}^2+\alpha_{i}, \
i=0,1,\ldots,p-1,
\label{pgf1}
\end{equation}
\begin{equation}
f_p(\mathbf{s})=\alpha_p+\delta_{p}s_{p}.
\label{pgf2}
\end{equation}
It is clear that the $i-th$ component of $\mathbf{f}(\mathbf{s})$ expresses all the possible outcomes of a cell of type $i$  in one time step. Letting $\mathbf{1}=[1,\ldots,1]$,  this concept is  expressed by the property that $f_i(\mathbf{1}) = 1$, which implies that $\alpha_i+\delta_i+\gamma_i=1, \ i=0,1,2,\ldots, p-1$ and, because we assume that cells of generation $p$ can only die or survive,  $\alpha_p+\delta_p=1$. This assumption is due to the fact that in our experimental setup, the CFSE fluorescence intensity becomes negligible for generations higher than $p$.

The moments of the MGW process can be expressed in terms of derivatives of
$f_i(s)$ (see ({\it 1, 2\/})). Specifically, the mean value of type $j$ cell count at the time point $1$ starting from a single cell of type $i$ at the time point $0$, is given by
\begin{equation}
\tilde{M}_{dr,ij}= E[{Z}_{dr,j}(1)\mid\mathbf{Z}_{dr}(0)=\mathbf{e}_{i}]=\frac{ \partial{f_{i}(\mathbf{s})}}{ \partial{s_{j}}}\bigg|_{\mathbf{s}=\mathbf{1}} \, i,\ j=0, 1,\ldots, p , \
\label{meanij}
\end{equation}
where $\mathbf{e}_i, 0 \le i \le  p$ denotes a vector whose $i+1-th$ component is $1$ and whose other components are $0$. Since our Markov process is stationary, we can easily derive the transition matrix $\tilde{\mathbf{M}}_{dr}\in \mathbb{R}^{(p+1, p+1)}$ which maps  cell counts at time  $n$ to cell counts at time $n+1$:
\begin{equation}
{\tilde{\mathbf{M}}_{dr}} = \left(\begin{array}{cccccc}
 \delta_{0}& 2\gamma_{0}& 0& 0& 0&0 \\
    0 & \delta_{1}&2\gamma_{1}&0&\ldots & \ldots \\
    0 &     0& \delta_{2}&2\gamma_{2}& 0& 0\\
    \vdots & \vdots & 0& \ddots & \ddots & 0\\
    \vdots &\vdots &\vdots &0 &\delta_{p-1}& 2\gamma_{p-1}\\
0& 0& 0& 0&0& \delta_{p}
\end{array}
\right) \ .
\end{equation}
It is easy to check that the mean value of $\mathbf { Z}_{dr}(n)$ conditional to a given  state $\mathbf { Z}_{dr}(n-1)$ is:
\begin{equation}
\label{meandr}
 E[\mathbf{Z}_{dr}(n) \mid \mathbf { Z}_{dr}(n-1)] = \mathbf{Z}_{dr}(n-1) \tilde{\mathbf{M}}_{dr} \ .
\end{equation}
The mean value $\boldsymbol{\mu}_{dr}$ is obtained by taking expectaction with respect to $\mathbf { Z}_{dr}(n-1)$ in \eqref{meandr}:
\begin{equation}
\label{meandr1}
 \boldsymbol{\mu}_{dr}(n)=[\mu_{dr,0}(n),\ldots,\mu_{dr,p}(n)]=E[\mathbf{Z}_{dr}(n)] = \boldsymbol{\mu}_{dr}(n-1) \tilde{\mathbf{M}}_{dr} \ .
\end{equation}
For computing the covariance matrix  of cell counts $\mathbf{Z}_{dr}(n)$, we start by calculating the conditional second order moment  $E[\mathbf{Z}^T_{dr}(n)\mathbf{Z}_{dr}(n) \mid \mathbf { Z}_{dr}(n-1)]$:

\begin{equation}
\begin{array}{l}
E[\mathbf{Z}^T_{dr}(n)\mathbf{Z}_{dr}(n) \mid \mathbf { Z}_{dr}(n-1)] =  \\
E[\mathbf{Z}^T_{dr}(n)\mid \mathbf{Z}_{dr}(n-1)] E[\mathbf{Z}_{dr}(n)\mid \mathbf{Z}_{dr}(n-1)]+ Cov[\mathbf{Z}_{dr}(n) \mid \mathbf{Z}_{dr}(n-1)]= \\
\tilde{\mathbf{M}}^T_{dr}\mathbf{Z}^T_{dr}(n-1)\mathbf{Z}_{dr}(n-1)\tilde{\mathbf{M}}_{dr}+\sum_{l=0}^{p}\mathbf{V}_lZ_{dr,l}(n-1) \ ,  \end{array}
\label{condmom}
\end{equation}
\noindent
where $\mathbf{V}_l$ represents the one step covariance matrix  for one cell present in the state $\mathbf{Z}(0)=\mathbf{e}_l$ (see ({\it 1\/})):
\begin{equation}
(\mathbf{V}_l)_{ij} = \left[\frac{\partial^{2} f_{l}(\mathbf{s})}{\partial s_{i}\partial s_{j}}-\frac{\partial f_{l}(\mathbf{s})}{\partial s_i}\frac{\partial f_{l}(\mathbf{s})}{\partial s_j}\right]_{\mathbf{s}=\mathbf{1}}, \ \ i \neq j
\label{vij}
\end{equation}
\begin{equation}
(\mathbf{V}_l)_{ii} = \left[\frac{\partial^2 f_{l}(\mathbf{s})}{\partial^{2} s_i}+\frac{\partial f_l(\mathbf{s})}{\partial s_{i}}\left( 1-\frac{\partial f_l(\mathbf{s})}{\partial s_{i}}\right)\right]_{\mathbf{s}=\mathbf{1}}.
\label{vii}
\end{equation}
By taking expectation with respect to $\mathbf {Z}_{dr}(n-1)$ in \eqref{condmom}  we get the second order moment $\mathbf{S}^*_{dr}(n)$:
\begin{equation}
\label{sordmom}
\mathbf{S}^*_{dr}(n)=E[\mathbf{Z}^T_{dr}(n)\mathbf{Z}_{dr}(n)]=\tilde{\mathbf{M}}_{dr}^T\mathbf{S}^*_{dr}(n-1) \tilde{\mathbf{M}}_{dr} +\sum_{l=0}^{p}\mathbf{V}_l\mu_{dr,l}(n-1) \ .
\end{equation}
Then, we get the covariance matrix $\mathbf{S}_{dr}(n)$ recursion by considering that $\mathbf{S}_{dr}(n)=\mathbf{S}^*_{dr}(n)-\boldsymbol{\mu}^T_{dr}(n) \boldsymbol{\mu}_{dr}(n)$:
\begin{equation}
\label{covdr}
\mathbf{S}_{dr}(n) = \tilde{\mathbf{M}}_{dr}^T\mathbf{S}_{dr}(n-1) \tilde{\mathbf{M}}_{dr} +\sum_{l=0}^{p}\mathbf{V}_l\mu_{dr,l}(n-1) \ .
\end{equation}

Now, we want to extend this basic model by allowing the presence of migration from the draining lymph node.  To this purpose,
we introduce the r.v.$'$s $\eta_i$ $i=1,2,\ldots, p-1$ representing binomial random variables associated to the cells of different types.
We assume that $\eta_{i}=1$ if the cell of type $i$ migrates in a time step $\Delta t$ and  $\eta_{i}=0$ if the cell of type $i$ doesn't migrate. We denote by $m_i$ and $1-m_i, \ i=1,2,\ldots,p-1$ the probabilities of the two possible events, respectively. Notice that, we assume here that the \textit{naive} T cells (type $0$) and the highest generation cells(type $p$) have null migration probabilities, i.e., $m_0 = m_p=0$. This is in accordance with biological knowledge on the process and the fact that the CFSE measuring equipment has limited resolution.\\
Now, we introduce the pgf's describing the MGW process with the presence of migration:

\begin{equation}
\begin{array}{rcl}
f_{dr,i}(\mathbf{s})& =& \delta_{i} (1-m_i) s_{i}+\gamma_{i} (1-m_{i+1}) s_{i+1}^2+\alpha_{i}, \ i=0,1,\ldots,p-1 \\
f_{dr,p}(\mathbf{s})&=&\delta_{p} s_{p} + \alpha_p \ .
\end{array}
\label{eq:fdrmig}
\end{equation}
Notice that the parameters $\alpha_i, i=0, 1, \ldots, p-1$ in the first equation above have a meaning different from the `mortality'  rate of the basic model (without migration). Here, they take into account the cumulative effect of death and migration of cells of the various generations.

Hence, we can use \eqref{meanij} with the pgf's in \eqref{eq:fdrmig} for computing the transition matrix of the process in our source node $\mathbf{M}_{dr} \in \mathbb{R}^{p+1,p+1}$ in the presence of migration:

\begin{equation}
{\mathbf{M}_{dr}} = \left(\begin{array}{cccccc}
 \delta_{0}& 2\gamma_{0}(1-m_1)& 0& 0& 0&0 \\
    0 & \delta_{1}(1-m_1)&2\gamma_{1}(1-m_2)&0&\ldots & \ldots \\
    0 &     0& \delta_{2}(1-m_2)&2\gamma_{2}(1-m_3)& 0& 0\\
    \vdots & \vdots &\vdots& \ddots&\ddots  & 0\\
    \vdots &\vdots &\vdots &0 &(1-m_{p-1})\delta_{p-1}& 2\gamma_{p-1}\\
0& 0& 0& 0&0& \delta_{p}
\end{array}
\right) \ .
\end{equation}\

 Then, with a slight abuse of notation, we will replace the mean value and covariance recursive equations \eqref{meandr1} and  \eqref{covdr} with the following expressions holding in the presence of migration:
\begin{equation}
\boldsymbol{\mu}_{dr}(n)=\boldsymbol{\mu}_{dr}(0)\mathbf{M}_{dr}^n \ ,
\label{meanmig}
\end{equation}

\begin{equation}
\mathbf{S}_{dr}(n)=\mathbf{M}_{dr}^T\mathbf{S}_{dr}(n-1)\mathbf{M}_{dr}+\sum_{l=0}^{p}\mathbf{V}_l\mu_{dr,l}(n-1) \ ,
\label{covmig}
\end{equation}
where the one time step covariance matrices $V_l$ are again computed through \eqref{vij} and \eqref{vii}.

Now, we can derive the conditional expected value of the migrating T cells from the draining lymph node by introducing the probability generating function for one cell of type $l$  in one time step:
\begin{equation}
\begin{array}{rcl}
f_{mig,0}(\mathbf{s})&=&\gamma_0m_1s^2_1 \\
f_{mig,l}(\mathbf{s})&=&\delta_lm_ls_l+\gamma_lm_{l+1}s^2_{l+1} \ , l=1, \ldots, p-2 \ ,\\
f_{mig,p-1}(\mathbf{s})&=&\delta_{p-1}m_{p-1}s_{p-1}\\
 f_{mig,p}&=& 0 \
 \end{array}
\label{eq:fmig}
\end{equation}
$$
\mathbf{s}=[s_{0}, s_{1}, s_{2},\dots, s_{p}] \in \mathbb{C}^{p+1}, \vert s_l \vert \leq 1 \ , l= 0, 1, \ldots, p \ .$$
Then, we get the conditional expected value of migrating T cells at time point $n$ as:
\begin{equation}
\label{condzmig}
E[\mathbf{Z}_{mig}(n)|\mathbf{Z}_{dr}(n-1)]=\mathbf{Z}_{dr}(n-1)\mathbf{M}_{mig}
\end{equation}
where again the transition matrix $\mathbf{M}_{mig} \in \mathbb{R}^{(p+1,p+1)}$ is computed through \eqref{meanij} with the pgf's in \eqref{eq:fmig}:
\begin{equation}
{\mathbf{M}_{mig}} = \left(\begin{array}{cccccc}
 0& 2\gamma_0m_1& 0& 0& 0&0 \\
    0 &m_1 \delta_{1}&2\gamma_1m_2&0&\ldots & \ldots \\
    0 &     0&m_2 \delta_{2}&2\gamma_2m_3& 0& 0\\
    \vdots & 0 &\vdots& \ddots&\ddots  & 0\\
    \vdots &\vdots &\vdots &0 &m_{p-1}\delta_{p-1}& 0\\
0& 0& 0& 0&0& 0
\end{array}
\right) \ .
\end{equation}\

By taking expectation over $\mathbf{Z}_{dr}(n-1)$ in \eqref{condzmig}, we get the mean value of migrating T cells $\boldsymbol{\mu}_{mig}(n) \in \mathbb{R}^{p+1}$ at time point $n$:
\begin{equation*}
\boldsymbol{\mu}_{mig}(n)=\boldsymbol{\mu}_{dr}(n-1)\mathbf{M}_{mig} \ .
\end{equation*}
To compute the covariance matrix at time point $n$, $\mathbf{S}_{mig}(n) \in \mathbb{R}^{(p+1,p+1)}$, we first derive the conditional and unconditional second order moments of the migrating cell counts:
\begin{equation}
\begin{array}{l}
E[\mathbf{Z}^T_{mig}(n)\mathbf{Z}_{mig}(n) \mid \mathbf{Z}_{dr}(n-1)]  = \\
 E[\mathbf{Z}^T_{mig}(n) \mid \mathbf{Z}_{dr}(n-1)]E[\mathbf{Z}_{mig}(n) \mid \mathbf{Z}_{dr}(n-1)]+ Cov[\mathbf{Z}_{mig}(n)\mid \mathbf{Z}_{dr}(n-1)] = \\
  \mathbf{M}^T_{mig}\mathbf{Z}^T_{dr}(n-1)\mathbf{Z}_{dr}(n-1)\mathbf{M}_{mig}+\sum_{l=0}^{p}\mathbf{V}_{mig,l} Z_{dr,l}(n-1) \ ,
\end{array}
\end{equation}
where $\mathbf{V}_{mig,l}$ represents the one step covariance matrix of one cell of type $l$ available for migration from the draining lymph node. These matrices  can be computed again through equations \eqref{vij} and \eqref{vii}, with the pgf's in \eqref{eq:fmig}.

By taking expectation over $\mathbf{Z}_{dr}(n-1)$, we get:
\begin{equation*}
\mathbf{S}^*_{mig}(n)=E[\mathbf{Z}^T_{mig}(n)\mathbf{Z}_{mig}(n)]=\mathbf{M}_{mig}^T\mathbf{S}^*_{dr}(n-1)\mathbf{M}_{mig}+\sum_{l=0}^{p}\mathbf{V}_{mig,l} \mu_{dr,l}(n-1).
\end{equation*}
Then, the covariance matrix is readily obtained:
\begin{equation}
\begin{array}{rcl}
\mathbf{S}_{mig}(n)& =& \mathbf{S}^*_{mig}(n)-\boldsymbol{\mu}^T_{mig}(n)\boldsymbol{\mu}_{mig}(n)\\
& = &\mathbf{M}_{mig}^T\mathbf{S}_{dr}(n-1)\mathbf{M}_{mig}+\sum_{l=0}^{p}\mathbf{V}_{mig,l} \mu_{dr,l}(n-1) \ .
\end{array}
\label{Smig}
\end{equation}

\subsection*{Transfer compartment}
\begin{figure}[htbp]
	\centering
	\includegraphics[scale=0.5]{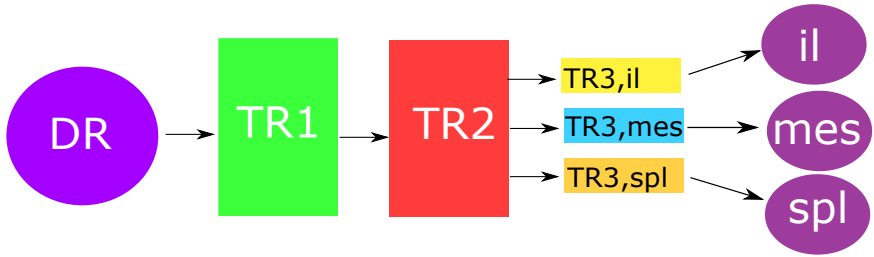}
	\caption{\textbf{Network model schematization}. We assume our network model composed by draining lymph node(dr), transfer compartment with its subcompartments TR1, TR2 and TR3 where the T cell immigrate in the mesenteric (mes), iliac (il) distal lymph nodes and spleen (spl).}
	\label{fig:figure1}
\end{figure}

Assuming a time step  equal to 4 hours and that T cells take approximately 12 hours to move from the draining lymph node to the distal compartments, the transfer compartment can be built by assembling 3 serial subcompartments TR1, TR2 and TR3 (see Fig. \ref{fig:figure1}).
 Of course, any different choice would not impact on the structure of the model, which is very flexible in this respect. Clearly, a rough knowledge of the migration time and of the time step, i.e., the mean division time, is necessary for appropriately structuring the transfer compartment.

We assume that migrating T cells $\mathbf{Z}_{mig}(n)$ at time point $n$ are  located in the TR1 subcompartment.
 We denote by $\mathbf{Z}_{TR1}(n)$ the T cells counts vector of the first subcompartment. Then, the mean value $\boldsymbol{\mu}_{TR1}(n)\in \mathbb{R}^ {p+1}$ and covariance matrix $\mathbf{S}_{TR1}(n)  \in \mathbb{R}^{(p+1,p+1)}$ are:
\begin{eqnarray}
\boldsymbol{\mu}_{TR1}(n)&=&E[\mathbf{Z}_{TR1}(n)]=\boldsymbol{\mu}_{mig}(n)\\
\mathbf{S}_{TR1}(n)&=&\mathbf{S}_{mig}(n) \ .
\end{eqnarray}
After one time step in the TR1 compartment, the cells $\mathbf{Z}_{TR1}(n)$ undergo a transition while moving to the TR2 subcompartment, according to the pgf $\mathbf{f}_{tr}(\mathbf{s})= [f_{tr,0}, f_{tr,1}, \ldots, f_{tr,p}]$, defined as:
\begin{equation}
\begin{array}{rcl}
f_{tr,0}(\mathbf{s})&=&0 \\
f_{tr,i}(\mathbf{s})&=& f_{i}(\mathbf{s}), \  i=1,2,\ldots,p-1 \\
f_{tr,p}(\mathbf{s})&=& 0 \ .
\end{array}
\label{pgftrcomp}
\end{equation}
From these pgf's we easily derive the transition matrix $\mathbf{M}_{tr} \in \mathbb{R}^{(p+1,p+1)}$:

\begin{equation*}
{\mathbf{M}_{tr}} = \left(\begin{array}{cccccc}
 0& 0& 0& 0& 0&0 \\
    0 & \delta_{1}&2\gamma_1&0&\ldots & \ldots \\
    0 &     0&\delta_{2}&2\gamma_2& 0& 0\\
    \vdots & 0 &\vdots& \ddots&\ddots  & 0\\
    \vdots &\vdots &\vdots &0 &\delta_{p-1}& 0\\
0& 0& 0& 0&0& 0
\end{array}
\right) \ .
\end{equation*}\
Hence the conditional mean value of cell counts in the TR2 compartment is given by:
\begin{equation}
\boldsymbol{Z}_{TR2}(n)=E[\mathbf{Z}_{TR2}(n)|\mathbf{Z}_{TR1}(n-1)]=\mathbf{Z}_{TR1}(n-1)\boldsymbol{M}_{tr} \ .
\label{Ztr2}
\end{equation}
By taking expectation  over $\mathbf{Z}_{TR1}(n-1)$ in \eqref{Ztr2}, we get:
\begin{equation}
\boldsymbol{\mu}_{TR2}(n)=E[\mathbf{Z}_{TR2}(n)]=\boldsymbol{\mu}_{TR1}(n-1)\boldsymbol{M}_{tr} \ .
\label{mutr2}
\end{equation}
The covariance matrix  $\mathbf{S}_{TR2}(n) \in \mathbb{R}^{(p+1,p+1)}$ at time point $n$  can be easily computed following the same steps as  for $\mathbf{S}_{mig}$:

\begin{equation}
\mathbf{S}_{TR2}(n) =\mathbf{M}^{T}_{tr}\mathbf{S}_{TR1}(n-1)\mathbf{M}_{tr}+
\sum_{l=1}^{p}\mathbf{V}_{tr,l}\mu_{TR1,l}(n-1) \ ,
\label{covtr2}
\end{equation}
where $\mathbf{V}_{tr,l}$ is obtained through \eqref{vij} and \eqref{vii} by using the pgf's introduced in \eqref{pgftrcomp}.

Concerning the splitting of the cell population among the the distal lymph nodes and spleen, we model this phenomenon as contemporary to the transition to the third transfer subcompartment. Hence, at the the end of the third transfer time step, we are able to compute the three distinct subpopulations  in the subcompartment TR3 (see Fig. 2). This step is described in the next subsection together with the proliferation in the distal nodes.

\subsection*{Distal lymph nodes and spleen}
The cell population evolution in the distal lymph nodes and spleen depends  on the intrinsic proliferation process and the immigration flow from the transfer compartment. Since we assume that splitting of the migrating population takes place during the transition from TR2 to TR3, we can compute the three flows of cells directed to the distal compartments, directly at the level of the third transfer subcompartment. The flow from the TR2 compartment towards the distal nodes is divided into three distinct components named $\mathbf{Z}_{TR3,spl}(n),\mathbf{Z}_{TR3,il}(n),\mathbf{Z}_{TR3,mes}(n)$  to identify the number of cells addressed to the spleen,   iliac and mesenteric lymph nodes, respectively.   We will denote by $\rho_{spl},\rho_{mes}, \rho_{il}, \ (\rho_{il}+\rho_{mes}+ \rho_{spl}=1)$ the probabilities that a cell decides to move to the spleen, the mesenteric or the iliac node. At the same time, our model takes care of a proliferation transition during the `splitting' time step, according to the  MGW pgf modified to account for the splitting. In the remaining part of this section, we will make reference to the spleen, giving for granted that similar formulas hold for the other distal compartments. The transition function from $\mathbf{Z}_{TR2}(n)$ to  $\mathbf{Z}_{TR3,spl}(n+1)$ is regulated by the following pgf:

\begin{equation}
\mathbf{f}_{TR3,spl}(\mathbf{s})=[f_{TR3spl,0}(\mathbf{s}),f_{TR3spl,1}(\mathbf{s}), f_{TR3spl,2}(\mathbf{s}),\ldots, f_{TR3spl,p}(\mathbf{s})]^T
\label{pgfspleen}
\end{equation}
where
$$
\mathbf{s}=[s_{0}, s_{1}, s_{2},\dots, s_{p}] \in \mathbb{C}^{p+1}, \vert s_i \vert \leq 1 \ ,$$
and
\begin{equation}
\begin{array}{rcl}
f_{TR3spl,0}(s)&=&0\\
f_{TR3spl,i}(s)&=& \rho_{spl}\delta_{i}s_{i}+\rho_{spl}\gamma_{i}s_{i+1}^2+(1-\rho_{spl}\delta_{i}-\rho_{spl}\gamma_{i}),
i=1,\ldots,p-1\\
f_{TR3spl,p}(s)&=&(1-\rho_{spl}\delta_{p})+\rho_{spl}\delta_{p}s_{p} \ .
\end{array}
\label{pgfspleen1}
\end{equation}
 Then, the conditional mean value of $\mathbf{Z}_{TR3,spl}(n)$ is given by:
\begin{equation}
\mathbf{Z}_{TR3,spl}(n) = E[\mathbf{Z}_{TR3,spl}(n)|\mathbf{Z}_{TR2}(n-1)]=\mathbf{Z}_{TR2}(n-1)\mathbf{M}_{split,spl}
\end{equation}

where

\begin{equation}
\mathbf{M}_{split,spl}=\rho_{spl}\mathbf{M}_{tr} \ .
\end{equation}

Expectation over $\mathbf{Z}_{TR2}(n-1)$ leads to:
\begin{equation}
\boldsymbol{\mu}_{TR3,spl}(n)=\boldsymbol{\mu}_{TR2}(n-1)\mathbf{M}_{split,spl}
\end{equation}
By using the same machinery adopted before, we get the recursion for the cell counts covariance matrix:
\begin{equation}
\mathbf{S}_{TR3,spl}(n)=\mathbf{M}^{T}_{split,spl}\mathbf{S}_{TR2}(n-1)\mathbf{M}_{split,spl} +\sum_{l=1}^{p}\mathbf{V}_{(split,spl),l}\mu_{TR2,l}(n-1)\ ,
\label{immcov}
\end{equation}
where $\mathbf{V}_{(split,spl),l}$ the one step covariance matrix for one cell of type $l$, which can be derived by exploiting \eqref{pgfspleen1}.

Now, we turn to the computation of the mean and the covariance matrix of cell counts in the spleen. Before doing this, we have to quantify the immigrating flow $\boldsymbol{\lambda}_{spl}(n)$ in the spleen and its covariance matrix:
\begin{equation}
\begin{array}{rcl}
\boldsymbol{\lambda}_{spl}(n)=E[\mathbf{Z}_{spl}(n)|\mathbf{Z}_{spl}(n-1)=0]&=& \boldsymbol{\mu}_{TR3,spl}(n)\\
\mathbf{W}_{spl}(n) =Cov[\mathbf{Z}_{spl}(n)|\mathbf{Z}_{spl}(n-1)=0]&=&\mathbf{S}_{TR3,spl}(n) \ .
\end{array}
\label{immflow}
\end{equation}
Now, we can compute the conditional expected value of cell counts in the spleen, by considering the MGW process with the immigration flow ({\it 40\/}):

\begin{equation}
\mathbf{Z}_{spl}(n)=E[\mathbf{Z}_{spl}(n)|\mathbf{Z}_{spl}(n-1)]=\mathbf{Z}_{spl}(n-1)\mathbf{M}_{dist}+\boldsymbol{\lambda}_{spl}(n) \ ,
\end{equation}
where $\mathbf{M}_{dist}=\mathbf{M}_{tr}$.
By taking expectation over $\mathbf{Z}_{spl}(n-1)$, we get:

\begin{equation}
\boldsymbol{\mu}_{spl}(n)=\boldsymbol{\mu}_{spl}(n-1)\mathbf{M}_{dist}+\boldsymbol{\lambda}_{spl}(n) \ .
\end{equation}

Finally, we can compute the covariance matrix $\mathbf{S}_{spl}(n)$ through the usual machinery, by suitably taking into account the immigrating population adding to the standard MGW process:
\begin{equation}
\mathbf{S}_{spl}(n)=\mathbf{M}^T_{dist}\mathbf{S}_{spl}(n-1)\mathbf{M}_{dist}+\sum_{l=1}^{p}\mathbf{V}_{tr,l}\mu_{spl,l}(n-1)+\mathbf{W}_{spl}(n)\ ,
\label{covspleen}
\end{equation}
where $\mathbf{W}_{spl}(n)$ is given by \eqref{immflow} and \eqref{immcov}.
\section{Normal approximation for the log likelihood function of the cell relative frequencies}
All \textit{in vivo} CFSE experiments require to sacrifice animals to collect data. This means that measurements taken at different time points actually refer to different individuals. This fact introduces an  inter-individual stochastic variability in terms of T cell counts which need be addressed in our inference scheme.  \\
To take into account this issue, model identification has been performed by using \textit{relative frequencies} as model variables instead of T cell counts. This choice mainly based on the results in ({\it 3 \/}), allows one to derive an explicit expression for the asymptotic log-likelihood function of the relative frequencies when dealing only with one draining lymph node({\it 3--5\/}). When dealing with the entire network, the inference problem becomes much more complicated. Also in our context, we take a relative frequency approach, which will allow us to derive a closed form expression for a normal approximation to the log-likelihood function. To this purpose, we introduce the  vectors  $\mathbf{Z}(n)=[\mathbf{Z}_{dr}(n),\mathbf{Z}_{spl}^*(n),\mathbf{Z}_{il}^*(n),\mathbf{Z}_{mes}^*(n)] \in \mathbb{R}^{p_{tot}}$ and $\boldsymbol{\mu}(n)=[\boldsymbol{\mu}_{dr},\boldsymbol{\mu}_{spl}^*(n),\boldsymbol{\mu}_{il}^*(n),\boldsymbol{\mu}_{mes}^*(n)]\in \mathbb{R}^{p_{tot}}$ where $p_{tot}=4p+1$ and $\mathbf{Z}_{spl}^*(n),\mathbf{Z}_{il}^*(n),\mathbf{Z}_{mes}^*(n),\boldsymbol{\mu}_{spl}^*(n),\boldsymbol{\mu}_{il}^*(n),\boldsymbol{\mu}_{mes}^*(n) \in \mathbb{R}^{p}$  represent the cell counts and the mean values vectors computed in the previous section, with the exclusion of the first component representing  \textit{naive} (type $0$) T cells. Similarly, we introduce the covariance matrices $\mathbf{S}_{spl}^*(n), \mathbf{S}_{il}^*(n),\mathbf{S}_{mes}^*(n)$ and build up the overall cell counts covariance matrix as
\begin{equation}
\mathbf{S}(n)=\left(\begin{array}{cccc}

    \mathbf{S}_{dr}(n)&0&0& 0\\
     0 &\mathbf{S}_{spl}^*(n)&0&0 \\
      0& 0&\mathbf{S}_{il}^*(n)& 0\\
     0 & 0& 0 & \mathbf{S}_{mes}^*(n)

\end{array}
\right) \ \in \mathbb{R}^{(p_{tot}, p_{tot})}.
\end{equation}
We highlight that the  matrix $\mathbf{S}(n)$ is introduced  to incorporate  the measurements taken at the different lymph nodes and spleen in the likelihood function. The block diagonal form is justified by the fact that the  measurements performed in the different model compartments are independent. In order to construct the likelihood function,
let $\mathbf{\Delta}(n) \in \mathbb{R}^{p_{tot}}$ be the vector random variable representing the relative frequencies at time point $n$. The  $i-th$ component of $\mathbf{\Delta}(n)$ is defined as
\begin{equation}
\Delta_{i}(n)=\frac{Z_i(n)}{U(n)},
\end{equation}
 where $U(n)=\sum_{i=1}^{p_{tot}}Z_i(n)$. Notice that $\sum_{i=1}^{p_{tot}}\Delta_{i}(n)=1$. Let $\mathbf{r}(n)$ be a vector whose components are:
 \begin{equation}
 r_i(n)=\frac{\mu_{i}(n)}{\sum_{i=1}^{p_{tot}}\mu_{i}(n)} \ , i=1,\ldots, p_{tot} \ .
 \end{equation}
Following arguments similar to those used in ({\it 3\/}), let us introduce the matrix $\mathbf{A}(n)$ whose entries are given by:
 \begin{eqnarray}
 (A(n))_{ii}&=&(S(n))^{1/2}_{ii}(1-r_{i}(n)),\\
  (A(n))_{ij}&=&-(S(n))^{1/2}_{ii}(1-r_j(n)),\\
   i,j&=&1,2,\ldots,p_{tot}, \  i\neq j \nonumber .
 \end{eqnarray}
 Moreover, define
 \begin{equation}
\mathbf{D}(n)=\mathbf{A}(n)^{T}\mathbf{R}(n)\mathbf{A}(n),
 \label{D}
 \end{equation}
 where $\mathbf{R}(n)$ is the correlation matrix of the cell counts, whose entries are given by
 \begin{equation}
 (R(n))_{ij}=\frac{(S(n))_{ij}}{((S(n))^{1/2}_{ii}(S(n))^{1/2}_{jj}} \ .
 \end{equation}
Then, the mean and the covariance matrix of relative frequencies can be derived as:
 \begin{equation*}
 \begin{array}{rcl}
E[\mathbf{\Delta}(n)] &=& [ r_1(n), \ldots, r_{p_{tot}}(n)]^T \\
\boldsymbol{\Sigma}(n) &=& \mathbf{D}(n) / (\sum_{i=1}^{p_{tot}}\mu_i(n))^2 \ .
\end{array}
\label{normreduced}
 \end{equation*}
 Since we know that $\sum_{i=1}^{p_{tot}}\Delta_{i}(n)=1$,  the relative frequencies distribution is degenerate. To this purpose, considering any subset of  relative frequencies in the $p_{tot}$ dimensional vector of relative frequencies and the related covariance matrix, named (with abuse of notation) $\mathbf{\Delta}(n)=[\Delta_{1}(n),\ldots,\Delta_{(p_{tot}-1)}(n)]^T \in \mathbb{R}^{p_{tot}-1}$ and $\boldsymbol{\Sigma}(n) \in \mathbb{R}^{p_{tot}-1,p_{tot}-1}$, we get
 \begin{equation}
 \begin{array}{rcl}
E[\mathbf{\Delta}(n)] &=& [ r_1(n), \ldots, r_{(p_{tot}-1)}(n)]^T \\
\mathbf{\Sigma}(n) &=& \tilde{\mathbf{D}}(n) / (\sum_{i=1}^{p_{tot}}\mu_i(n))^2 \ ,
\end{array}
 \end{equation}
where $\tilde{\mathbf{D}}(n)$ is defined like in (\ref{D}) through suitably rearranged submatrices of $\mathbf{A}(n)$.
The above equations allow us to construct the normal approximation to the log-likelihood function for our estimation problem. In fact, assume that  $c_n$ mice are sacrificed at the time point $n$. Denote by $\boldsymbol{\zeta}_{k}(n)$, $k=1,2,\ldots,c_n$, the relative frequency measurements at the  timepoint $n$, and by $\mathbf{\Sigma}_{k}(n)$ the relative frequencies covariance matrix at the same time point. Then, the contribution to the normal approximation of the log likelihood function  at time  $n$ is given by:
\begin{equation}
\begin{split}
L(\boldsymbol{\theta};n)=\frac{-c_n}{2}\ln (2\pi)-\frac{1}{2}\sum_{k=1}^{c_n}\textit{log}(det(\mathbf{\Sigma}_{k}(n))- \frac{1}{2}\sum^{c_n}_{k=1}(\boldsymbol{\zeta}_{k}(n)-\mathbf{r}(n))^T (\mathbf{\Sigma}_k(n)^{-1}(\boldsymbol{\zeta}_{k}(n)-\mathbf{r}(n)) \ ,
\end{split}
\label{likelihood1}
\end{equation}
where $\mathbf{r}(n)$ and $\mathbf{\Sigma}_k(n)$ are functions of the probability parameter vector
$$\boldsymbol{\theta} = [\delta_0,\gamma_0,\delta_1,\gamma_1,\ldots,\delta_{p-1},\gamma_{p-1} ,\delta_p, m_1,\ldots,m_{p-1}, \rho_{spl}, \rho_{il}, \rho_{mes}]^T .$$

 Of course, if independent measurements are taken at different time points $n=n_1,\ldots,n_z$, then the global negative approximate log likelihood function is
 \begin{equation}
 {\cal L}(\boldsymbol{\theta}) = - \sum_{i=1}^{z} L(\boldsymbol{\theta}; n_i) \ .
 \label{likelihood}
 \end{equation}
Notice that actually the cost function defined in (\ref{likelihood1}) would represent the true log likelihood function if the relative frequencies were normally distributed: this is the case (asymptotically) when dealing with the draining lymph node only (see ({\it 3--5\/})). In our case, the cost  function (\ref{likelihood}) represents a normal approximation of the negative log likelihood function, because we have no guarantee on the asymptotic normality of relative frequencies. By minimizing the cost function function (\ref{likelihood}), we get parameter estimates:
 \begin{eqnarray}
 &\hat{\boldsymbol{\theta}}  = \arg{\{\min(\cal{L}(\boldsymbol{\theta})\}}\\ \nonumber
 &\text{s.t.} \quad \\ \nonumber
&0 \le \delta_i \le 1, \  0 \le \gamma_i \le 1, \ \delta_i+ \gamma_i \le 1, \ i=0,1,\ldots ,p-1\\ \nonumber
& 0 \le m_i \le 1, \ i=1,\ldots ,p-1\\ \nonumber
& 0 \le \delta_p \le 1,\\ \nonumber
& 0 \le \rho_{spl} \le 1, \ 0 \le \rho_{il} \le 1, \ 0 \le \rho_{mes} \le 1, \\ \nonumber
&\rho_{spl}+ \rho_{il}+\rho_{mes} = 1 \ .\\ \nonumber
 \label{maxlikelihood}
 \end{eqnarray}

\section*{References}
\begin{itemize}
\item[1.]  T. E. Harris, The theory of branching processes (2002).
\item[2.] Kimmel, D. E. Axelrod, Branching processes in biology. Interdisciplinary applied mathematics, (2002).
\item[3.] Yakovlev A.Y., Yanev N.M., Relative frequencies in multitype branching processes. The Annals of Applied Probability. 2009; p.p 1-14.
\item[4.] Pettini E, Prota G, Ciabattini A, Boianelli A, Fiorino F, Pozzi G,Vicino A., Medaglini D., Vaginal Immunization to Elicit Primary T-Cell Activation and Dissemination. PloS One. 2013;8(12):e80545.
\item[5.]Boianelli A, Pettini E, Prota G, Medaglini D, Vicino A. Identification of a branching process model for adaptive immune response. In: Proc. IEEE 52nd
Conference on Decision and Control (CDC), 2013. IEEE; 2013. pp. 7205-7210.
\end{itemize}
\end{document}